\documentclass[%
floatfix, reprint,
 amsmath,amssymb,
 aps,
]{revtex4-2}

\usepackage{graphicx}
\usepackage{dcolumn}
\usepackage{bm}

\usepackage{graphicx}
\usepackage{dcolumn}
\usepackage{bm}
\usepackage[utf8]{inputenc}
\usepackage[T1]{fontenc}
\usepackage{mathptmx}
\usepackage{etoolbox}
\usepackage{xcolor}
    \usepackage{multirow}%
    \usepackage{amsmath,amssymb,amsfonts}%
    \usepackage{amsthm}%
    \usepackage{mathrsfs}%
    \usepackage[title]{appendix}%
    \usepackage{textcomp}%
    \usepackage{booktabs}%
    \usepackage{algorithm}%
    \usepackage{algorithmicx}%
    \usepackage{algpseudocode}%
    \usepackage{listings}%
    
    \usepackage{relsize}
    \usepackage{upgreek}
    \usepackage[]{hyperref}
\begin{document}

\preprint{APS/123-QED}

 \title{Probing quantum mechanics using nanoparticle Schrödinger cats}

\author{Sebastian Pedalino}
\email{sebastian.pedalino@univie.ac.at} 
\affiliation{Faculty of Physics, University of Vienna, Boltzmanngasse 5, 1090 Vienna, Austria}
\affiliation{Vienna Doctoral School in Physics, University of Vienna, Boltzmanngasse 5, 1090 Vienna, Austria}
\author{Bruno E. Ramírez-Galindo}
\affiliation{Faculty of Physics, University of Vienna, Boltzmanngasse 5, 1090 Vienna, Austria}
\affiliation{Vienna Doctoral School in Physics, University of Vienna, Boltzmanngasse 5, 1090 Vienna, Austria}
\author{Richard Ferstl}
\affiliation{Faculty of Physics, University of Vienna, Boltzmanngasse 5, 1090 Vienna, Austria}
\affiliation{Vienna Doctoral School in Physics, University of Vienna, Boltzmanngasse 5, 1090 Vienna, Austria}
\author{Klaus Hornberger}
\affiliation{Faculty of Physics, University of Duisburg-Essen, Lotharstraße 1, 47048 Duisburg, Germany}
\author{Markus Arndt}
\email{markus.arndt@univie.ac.at}
\affiliation{Faculty of Physics, University of Vienna, Boltzmanngasse 5, 1090 Vienna, Austria}
\author{Stefan Gerlich}
\affiliation{Faculty of Physics, University of Vienna, Boltzmanngasse 5, 1090 Vienna, Austria}

\date{\today}

\begin{abstract}
The quantum superposition principle is a cornerstone of physics and at the heart of many quantum technologies. Yet, it is still often regarded counterintuitive because we do not observe its key features on the macroscopic scales of our daily lives. It is therefore intriguing to ask how quantum properties persist or change as we increase the size and complexity of objects. 
 A paradigmatic test for this question can be realized by matter-wave interferometry, where the motion of individual massive particles becomes delocalized and needs to be described by a wave function that spans regions far larger than the particle itself. Here we present an experimental platform extending matter-wave interference to a qualitatively new class of materials that can vary widely in mass and size. 
 We specifically demonstrate quantum interference of sodium nanoparticles, which can each contain more than 7'000~atoms at masses greater than 170'000~dalton. 
 They propagate in a Schrödinger cat state with a macroscopicity of $\mu$\,=\,15.5, surpassing all previous experiments by an order of magnitude and providing the most stringent exclusion limit for generic macrorealistic modifications of the Schrödinger equation to date.
    
\begin{description}
\item[Keywords]
Quantum superposition, Matter-wave interference, Quantum macroscopicity, Schrödinger Cat, Cluster science
\end{description}
\end{abstract}

\maketitle

When Louis de Broglie postulated that we need to `associate a periodic phenomenon with any isolated portion of matter or energy', he envisioned that these new ideas would `solve almost all the problems brought up by quanta' \cite{DeBroglie1923}. In fact, the quantum wave function has become a core concept of modern physics \cite{Schroedinger1926} and has stood up to all tests to date. However, it is still a matter of debate whether quantum physics is already the ultimate theory or if it needs to be extended to explain its transition into classical phenomena.  
This debate has sparked general interest in the scientific community, shown by a series of recent experiments that have pushed the limits of quantum mechanics. Single atoms were delocalized on the half-meter scale \cite{Kovachy2015}, or for times longer than a minute \cite{Panda2024}. Matter-wave interference was realized for massive macromolecules \cite{Fein2019} and a variety of mechanical oscillators were cooled to their quantum ground state. This includes nanomechanical cantilevers \cite{OConnell2010,Chan2011} as well as levitated nanoparticles \cite{Delic2020,Magrini2021,Tebbenjohanns2021,Piotrowski2023}. Recently, the vibration mode of a bulk acoustic resonator was prepared in a Schrödinger cat state delocalized to $10^{-18}$\,m, at an effective mass of 16\,$\mu$g \cite{Bild2023}, and in a quantum ground state with a mass of $494\,\mu$g\,\cite{Doeleman2023}. 

Here, we present our work on nanoparticle interferometry in a complementary regime. In our case, the center-of-mass position of clusters containing more than seven thousand atoms becomes delocalized over a distance exceeding the particle's diameter by more than an order of magnitude. This quantum state is analogous to Schrödinger's cat, a macroscopic quantum object that defies classical intuition \cite{Schroedinger1935}.

The unique combination of mass and delocalization is particularly well suited for probing theories that modify the Schrödinger equation through nonlinear and stochastic terms to suppress macroscopic superpositions~\cite{Bassi2013}.
Such macrorealistic models can be falsified by quantum superposition experiments as presented below.

\section{Experiment}\label{sec:MWI}
\begin{figure*}[t]
\center
\includegraphics[width=0.8\textwidth]{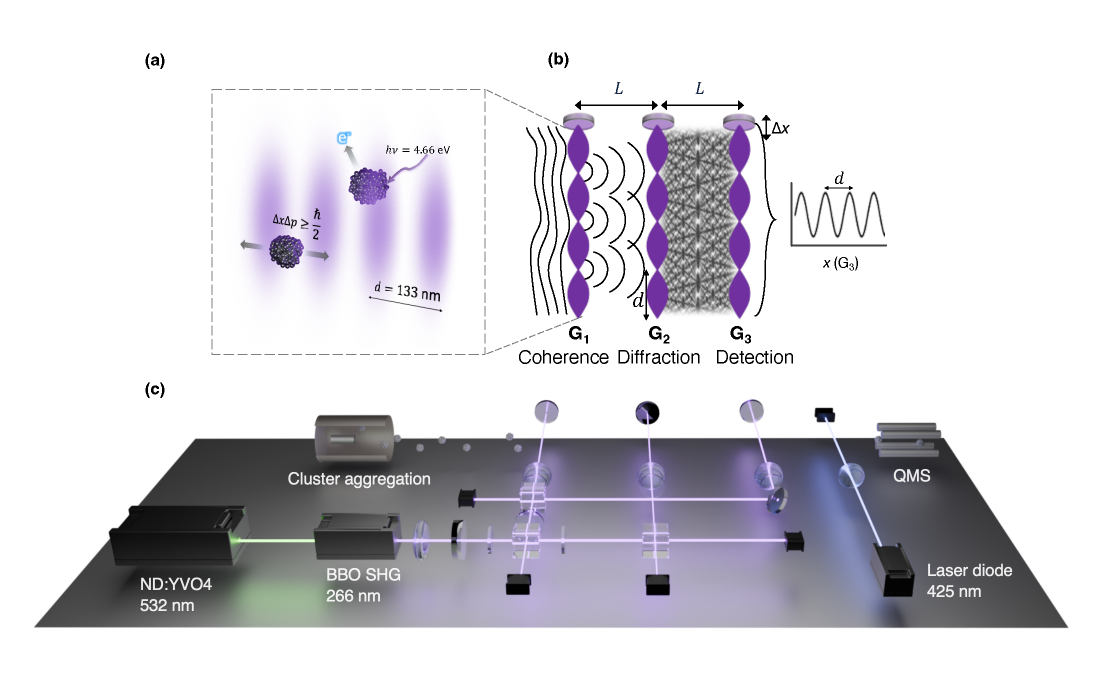}
\caption{\textbf{(a) Photo-ionizing gratings as beam splitters.} Clusters passing through the anti-nodes of the optical grating are ionized and removed, while those passing through the nodes remain neutral. This confines particles to a spatial region within the grating nodes resulting in a momentum uncertainty. The light field also induces a dipole moment, imprinting a position-dependent phase onto the clusters. \textbf{(b) Schematic of an optical Talbot-Lau interferometer}. Starting with incoherent matter-waves, the first grating (G$_1$) prepares coherence by spatially confining the particles, as described in (a). Transverse coherence grows towards G$_2$, behind which a Talbot-Lau carpet emerges in the near-field. Finally, a third grating acts as a position resolving detection mask scanned across the interference pattern. \textbf{(c) Schematic of the multi-scale cluster interference setup.} 
An effusive sodium source in an aggregation chamber generates the cluster beam. The beam is transmitted through several differential pumping stages into the interferometer chamber kept at ultra-high vacuum conditions ($\sim  9\times 10^{-9}$ \,mbar). The cluster beam overlaps with three perpendicular standing light waves equally spaced at a distance of $L=0.983$\,m, forming optical gratings with a period of $d=133$\,nm. The intensities of the first and third gratings are chosen such that they act as absorptive gratings, while the second grating is operated at lower laser intensity, realizing an optical phase grating. After passing through the interferometer, the remaining neutral clusters are photoionized using a 425\,nm laser diode and mass-filtered. The third grating is scanned transversely across the molecular beam. The integrated signal is then recorded as a function of the grating's displacement.}
\label{fig:setup}
\end{figure*}

The de Broglie wavelength $\lambda_\mathrm{dB}=h/mv$ of a matter wave beam is determined by Planck's constant $h$, the particle mass $m$, and its velocity $v$. Matter-wave interference at high masses requires both the preparation of low particle velocities and the ability to handle short de Broglie wavelengths.
In our multiscale cluster interference experiment (MUSCLE) we achieve this by combining a cryogenic metal cluster source with three ultraviolet diffraction gratings in a Talbot-Lau configuration, shown in Figure \ref{fig:setup}. 

Cluster aggregation sources enable scalable synthesis of particles across a wide mass range, and they are versatile in handling a variety of materials \cite{Haberland1991,deHeer1993,Pedalino2022}. Here, we prepare sodium clusters consisting of $5000 - 10000$ atoms, in a helium-argon mixture at 77\,K. They travel at velocities around  $160$\,m/s with de Broglie wavelengths between $10-22$\,fm. With a velocity spread of about $10$\,m/s, their longitudinal coherence length measures $160-350$\,fm. 

The short de Broglie wavelength makes far-field diffraction challenging even for grating periods on the hundred nanometer scale:
it would require collimating the beam to less than 100\,nm and observing diffraction at angles below 200\,nrad. However, 
in 1997, John Clauser proposed using near-field interferometry for grating-based coherent self-imaging of `small rocks and live viruses' \cite{Clauser1997}, noting that this approach is compact, tolerates initially incoherent beams, and offers high spatial resolution. This has been demonstrated with atoms \cite{Clauser1994,Fray2004}, x-rays \cite{Pfeiffer2006}, positrons \cite{Ariga2019}, as well as organic and tailored macromolecules \cite{Gerlich2007,Fein2019}. Here, we use it to open a window to matter-wave research with a whole new class of quantum objects, namely massive metal nanoparticles.

 A Talbot-Lau interferometer is built from three gratings with period $d$ and spacing close to the Talbot distance $L_T = d^2/\lambda_\text{dB}$ \cite{lau1948}. The first and third gratings act as periodic spatial filters to prepare matter-wave coherence in G$_1$ and to resolve the interference fringes that emerge at G$_3$. The second grating G$_2$ modulates the amplitude and phase of the cluster matter wave. Standing light waves are favored over nanomechanical diffraction gratings because their period is precisely defined, and their transmission amplitude can be modified in situ. 

\begin{figure*}[t]
    \centering
        \includegraphics[width=0.49\linewidth]{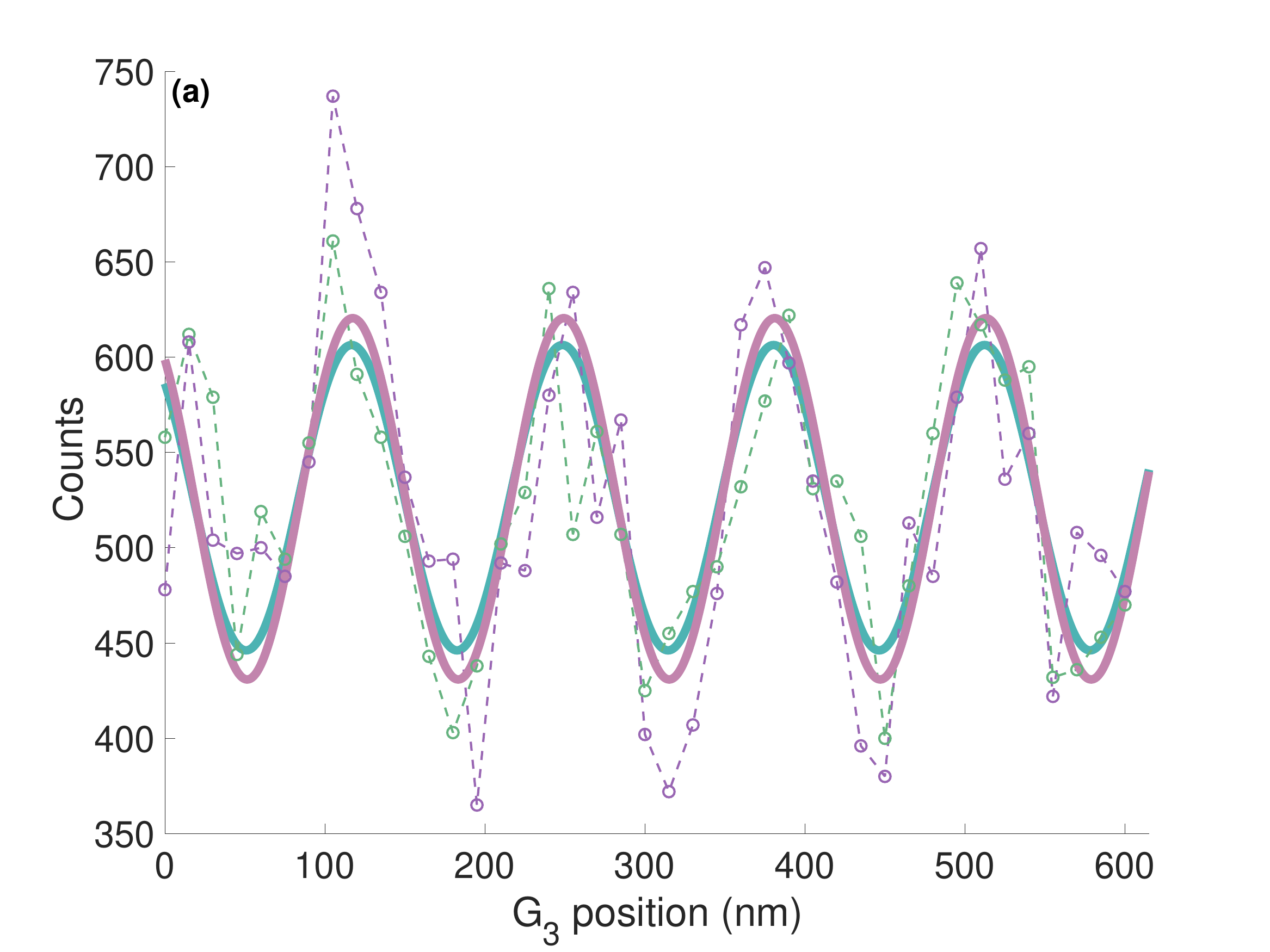}
        \includegraphics[width=0.49\linewidth]{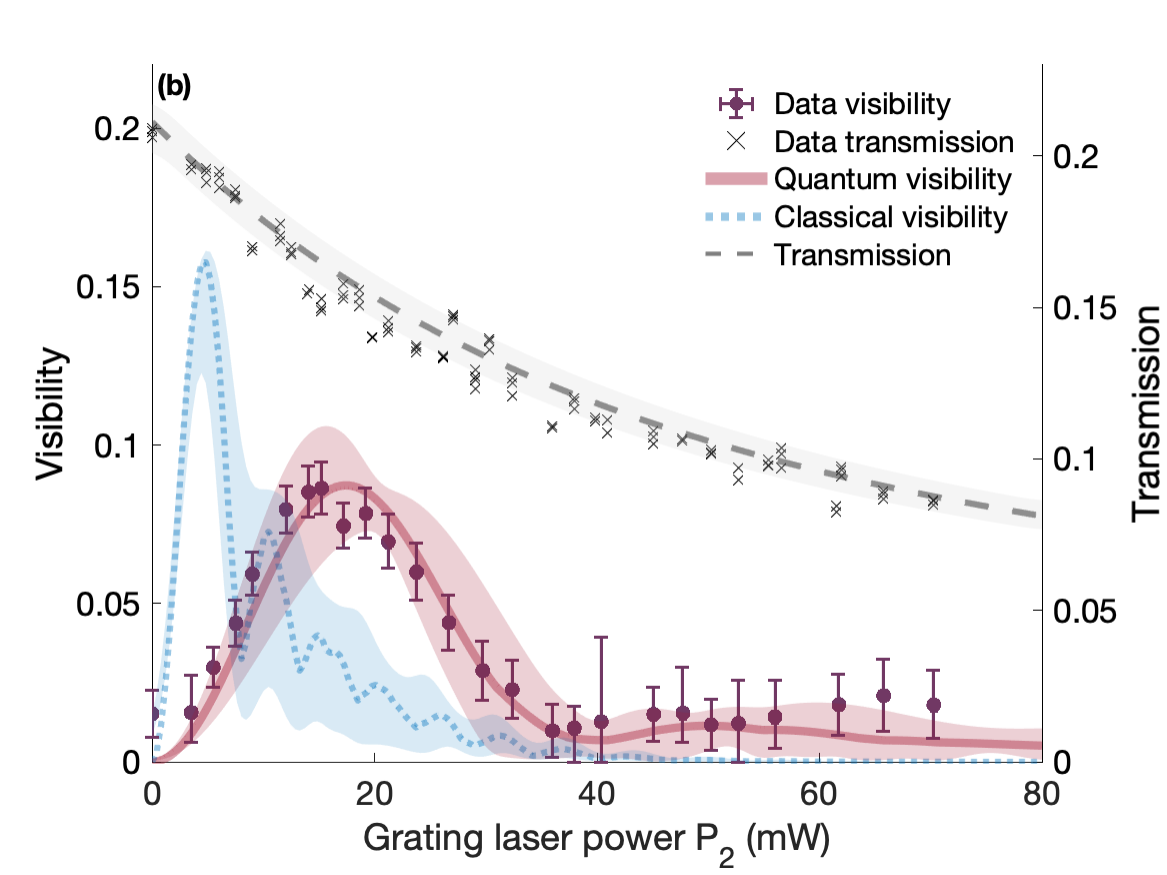}
\caption{
\textbf{(a) Interference fringes of sodium clusters with a mean mass of 172 kDa}. 
The experimental data of two independent measurement runs (purple and green dots) are fitted by a sine function (purple and green line) with a visibility of $V = 0.10 \pm 0.01$ and $V = 0.08 \pm 0.01$, for $P_{1} = (62 \pm 2) \, \text{mW}$, $P_{2} = (15.2 \pm 0.3) \, \text{mW}$ and $P_{3} = (68 \pm 2) \, \text{mW}$.
\textbf{(b) Fringe visibility versus grating laser power of G$_2$} 
Each data point represents the results of multiple independent interference scans of sodium clusters with masses centered around 172 kDa. Visibilities and error bars are derived from the sinusoidal fits and their corresponding $1\sigma$ confidence intervals. G$_{1,3}$ powers as above. 
The continuous red and the dashed blue lines depict the expected interference contrast according to the quantum and the classical model, respectively. The shaded areas display the uncertainties of the theory curve, based on the experimental $1\sigma$ limits of the molecular velocity, mass distribution, absorption cross section and optical polarizability. In this plot, both theory curves were scaled by the same global factor $0.78$.}
\label{fig:Interference}
\end{figure*}

In contrast to atom interferometry, where optical beam splitters are commonly tailored to specific electronic transitions \cite{Kasevich1991,Riehle1991}, ionization and phase gratings are compatible with a large variety of materials and particle sizes. Ultraviolet light serves well as an amplitude or photodepletion grating when the clusters in the antinodes are ionized and discarded. The standing light field additionally induces an oscillating dipole moment in the transmitted clusters, in proportion to their optical polarizability. Thus, it also imprints a spatially periodic phase shift onto the de Broglie wave associated with each nanoparticle.

The light for the three gratings is derived from a single-line green laser beam which is frequency doubled in an external cavity to produce up to 1\,W of power at 266\,nm. It is split into three partial beams, which are retro-reflected to form three standing light waves, separated by $0.983$\,m. 
Neutral clusters transmitted by the interferometer are photo-ionized and counted by a quadrupole mass spectrometer (QMS) using a conversion dynode and electron multiplier. 

We sample the interference patterns by scanning G$_3$ across the cluster beam while counting the number of transmitted clusters as a function of the G$_3$ position.
The resulting fringes are phase stable to within 3-5\,nm over several hours and can be fitted with a sinusoid to determine the visibility $V = (S_{\text{max}} - S_{\text{min}})/(S_{\text{max}} + S_{\text{min}})$ where the $S_{\text{max}}$ and  $S_{\text{min}}$ are the maximum and minimum of the fit, respectively.

\section{Results}\label{sec:Results}

In Figure~\ref{fig:Interference}a, we show two representative interference fringes of sodium clusters with a diameter around $8$\,nm and masses ranging from 143 to 197\,kDa. We have measured a fringe visibility of up to $V = 0.10 \pm 0.01$, which is limited by the finite photodepletion efficiency in the first and third gratings.

The observation of fringes in the cluster density distribution alone does not provide sufficient evidence for wave-like quantum propagation. They could also be explained by models in which the particles follow classical trajectories. In the presence of three nanomechanical gratings, classical flight-paths would produce moiré-like shadow patterns. A similar classical picture is conceivable for sinusoidal transmission gratings in G$_1$ and G$_3$ and a phase grating in G$_2$, where the latter acts as an array of microlenses due to the optical dipole force.   

To obtain clear evidence for the wave nature of the observed fringes, their visibility is analyzed as a function of the laser power $P_2$ of the second grating, shown by the solid circles in Figure \ref{fig:Interference}b. We compare this to the contrast predicted both by the classical (blue dotted line) and quantum model (solid red line). The quantum model is obtained by describing the matter-wave dynamics in phase space using the Wigner-Weyl formalism \cite{Nimmrichter2008}. It accounts for all coherent and incoherent grating interactions and enables a direct comparison with the prediction of classical mechanics (Suppl. Inf.).
\begin{figure*}[t]
    \centering
            \includegraphics[width=0.32\linewidth]{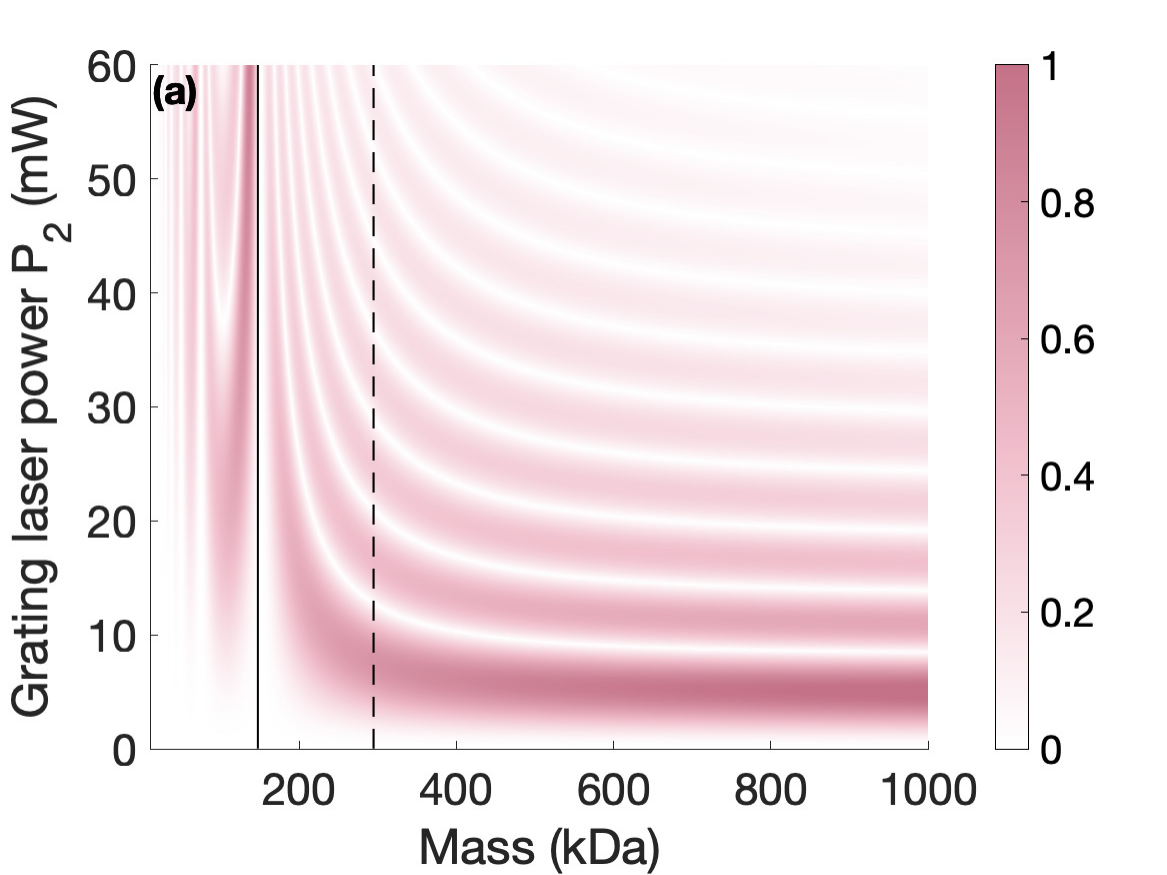}
            \includegraphics[width=0.32\linewidth]{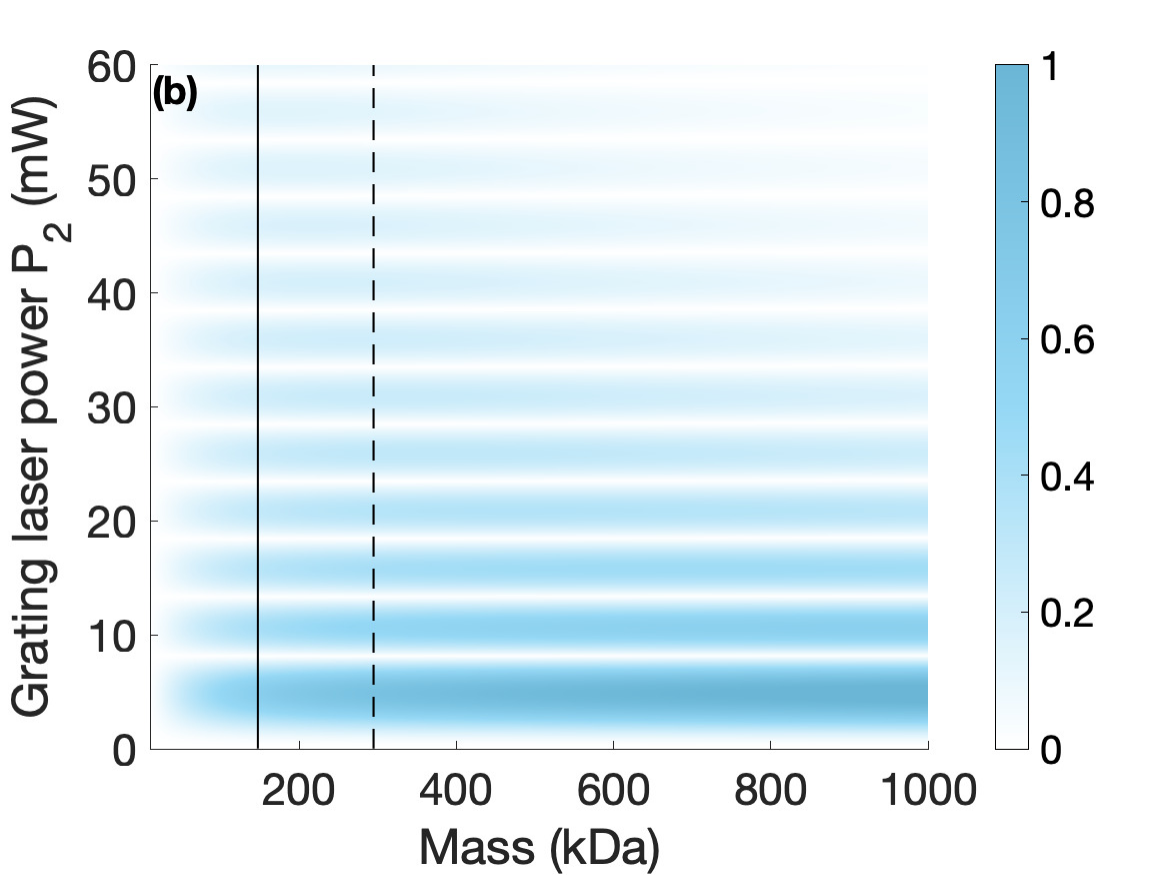}
            \includegraphics[width=0.32\linewidth]{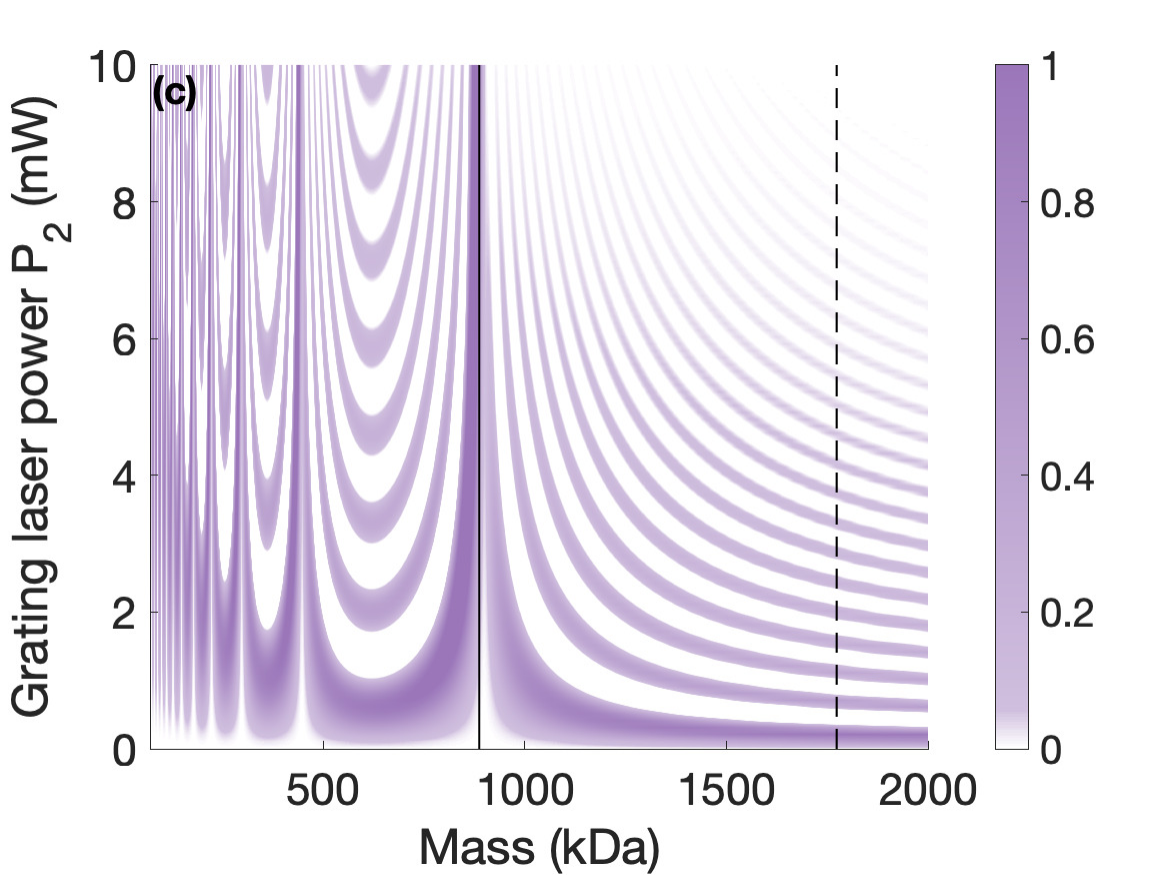}
\caption{
\textbf{Predicted fringe visibility as a function of cluster mass and G$_2$ laser power.} Results are shown for (a) the quantum model and (b) the classical model, which both include the effects of ionization and of the dipole force in the  grating interaction. Both calculations assume a mean velocity of $160$\,m/s, a Gaussian velocity spread of $10$\,m/s and grating powers of $P_1 = P_3 = 100$\, mW. The solid line marks the mass at which the Talbot length equals the interferometer length, while the dashed line indicates the mass for which half the Talbot length coincides with the interferometer length. The color scale indicates fringe visibilities $V$. For masses beyond the Talbot condition, the quantum and classical models converge. 
 (\textbf{c}) Slowing the particles to approximately $25$\,m/s will enable our setup to reliably distinguish quantum from classical dynamics for masses exceeding 1\,MDa.
}
 \label{fig:Visvsmass}
\end{figure*}
We account for the experimental constraints on velocity, ionization cross section, mass distribution, and polarizability by the shaded areas along the theory curves. 
Interferometer misalignment, gravitational and rotational phase averaging, mechanical vibrations, thermal or collisional decoherence can reduce the predicted contrast (Suppl. Inf.). We take this into account by a global scale factor of $0.78$ which is equally applied to the quantum and the classical prediction in this figure. With this single experimental factor included, our experiments are well described by the quantum model and clearly distinct from the classical prediction. 

Our assumptions regarding the clusters’ mass, size, and velocity distributions, as well as the mass dependence of their ionization cross section, are independently supported by the measured transmission probability as a function of the laser power in G$_2$ (dashed black curve). The model reproduces the experimental data (black crosses) very well, without any additional scaling factor.

For substantially more massive clusters, with masses between 400\,kDa and 1\,MDa, we observe even higher fringe visibilities of $V = 0.66 \pm 0.09$ (Suppl. Inf.). Although this may seem counterintuitive, it becomes plausible when we consider that the ionization cross section increases and the transmissive regions in each grating become narrower with increasing size of the cluster. 

However, the de Broglie wavelength in this mass range ($\lambda_\text{dB} \lesssim 3$\,fm) is too short to distinguish quantum from classical predictions, for our interferometer configuration (see Figure \ref{fig:Visvsmass}). For $L\leq L_T$, near-field matter-wave dynamics gradually transitions to geometrical optics, in agreement with Bohr's correspondence principle~\cite{bohr1928quantum}. 

Figure \ref{fig:Visvsmass} illustrates how the predicted visibilities from (a) quantum and (b) classical theory converge at high cluster masses. At the same time, it highlights a clear discrepancy between quantum and classical predictions in the mass range below 200\,kDa (see Figure \ref{fig:Interference}b). In panel (c) we show that it will become possible to unambiguously demonstrate the quantum wave nature of clusters in the MDa range if their velocities can be reduced to about 25\,m/s.

\section{Discussion}\label{sec:discussion}
Our experiments show that even a massive chunk of metal can be delocalized over far more than its own size, raising questions on the meaning of the objective reality of its position. The clusters are composite objects containing several thousand atoms, and similar to Schrödinger's cat - which cannot be described as either `dead' or `alive' - they cannot be assigned a unique trajectory. 
While this is conceptually the case for all matter-waves, from neutrons to molecules, our experiments advance this to a new level of macroscopicity.

To put our experiment into context with other demonstrations of quantum superposition states, we evaluate the macroscopicity measure $\mu$ as defined by \cite{Nimmrichter2013,Schrinski2019}.
This value quantifies to what extent a given quantum experiment probes the validity of quantum mechanics and how well it can exclude minimal modifications of the Schrödinger equation, which would break the quantum superposition principle at some macroscopic scale.

\begin{figure}[]
\center
\includegraphics[width=0.45\textwidth]{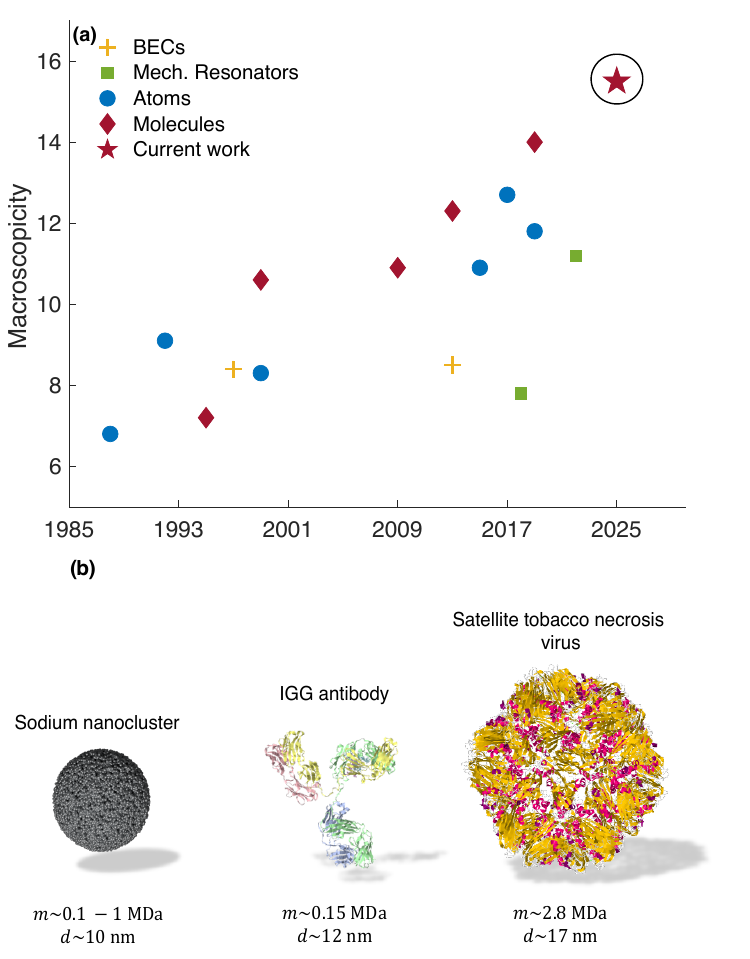}
\caption{\textbf{(a) Macroscopicity values of selected quantum experiments.} Blue circles: atom interferometry; red diamonds: molecule interferometry; orange crosses: Bose-Einstein condensates (BECs); green squares: mechanical resonators; red star: sodium nanoclusters in this study, with \textbf{$\mu = 15.5$}.
Reference data are taken from \cite{Nimmrichter2013,Schrinski2019,Schrinski2023} and explained in the Supplementary Information. \textbf{(b) Visualization of size and complexity:} The sodium clusters studied here behave as quantum particles at about $0.2$\,MDa and show high contrast up to the MDa regime. 
The number of atoms and their mass is  compatible with that of large proteins and small viruses (from protein database \cite{Rose2015}).}
\label{fig:Macro}
\end{figure}

Every successful demonstration of quantum interference falsifies a generic class of minimally invasive, macrorealistic modifications of quantum theory. To obtain the macroscopicity $\mu$, all raw experimental data are used to narrow down the parameter space of these models via Bayesian updating, as explained in \cite{Schrinski2019}. This requires a quantitative model for the outcome probabilities in the presence of macrorealistic modifications~\cite{Schrinski2020}. 
Any experimental imperfection and all decoherence processes are attributed to the macrorealistic modification and will therefore only decrease the macroscopicity (Methods).
From our data, we obtain the value $\mu=15.5$, which surpasses the previous record \cite{Schrinski2023} by an order of magnitude, as illustrated in Figure \ref{fig:Macro}a.

A major motivation for this line of research is to explore the quantum-classical interface bottom-up, systematically and with all parameters under control. Our interferometer is unique in that it can accept various metals and also dielectric nanoparticles with different mass densities in the same machine. An additional factor of one hundred in mass and in coherence time is still conceivable in a vertical interferometer~\cite{Kialka2022}; this would boost the attainable macroscopicity by six orders of magnitude in a ground-based experiment. This may open new opportunities to test the weak equivalence principle with vastly different types of matter, or to probe quantum physics in novel theories of gravity\,\cite{Oppenheim2023}.

On the applied side, coherent self-imaging creates a cluster density pattern in free flight, which can be shifted by external forces or directed momentum kicks. Particle-like properties, such as electric or magnetic susceptibility, can then be measured on clusters while they are propagating as delocalized waves. 
Such measurements are complementary to explorations in physical chemistry \cite{deHeer1993,xu2005,Merthe2016,Rivic2021, Antoine2002a} and promise higher force resolution.

The mass of our sodium clusters ($170$\,kDa) already surpasses that of a coconut cadang cadang viroid (CCCVd, $ 81$\,kDa\,\cite{Haseloff1982,Carabez2019}), or a protein such as immunoglobulin G (IgG, $ 150$\,kDa\,\cite{Rose2015}).  In the next generation of experiments, it is anticipated to approach the MDa mass range of small viruses, such as the STNV virus, sketched in Figure \ref{fig:Macro}b. 

Realizing quantum superpositions with massive bio-nanomaterials still demands significant advancements in beam preparation, coherent manipulation, and detection technologies. Recent progress in the generation \,\cite{Ekeberg2024,samanta2020,Mucha2022}, tools for coherent photodepletion \cite{Schaetti2019} and detection of beams of massive biomolecules \,\cite{Strauss2023} suggests that these challenges can be solved.

\section*{Methods}\label{sec:methods}

\subsection*{Quantum and classical model}

The theory of Talbot-Lau interference is best formulated in phase space using the Wigner-Weyl representation of quantum mechanics \cite{Nimmrichter2008}.
This framework can account for incoherent particle sources, phase and absorption gratings, and all laser-induced photophysical effects, as well as any relevant decoherence process. It also allows for a direct comparison between the predictions of quantum and classical mechanics within the same formalism and set of assumptions. 

For a cluster with mass $m$ and longitudinal velocity $v_z$, the probability of being detected behind the interferometer can be written as a Fourier series in the transverse position $x_3$ of G$_3$:  
\begin{eqnarray}
    S(x_3) &=& \sum_{\ell=-\infty}^\infty S_\ell \exp{\Big(i\frac{2\pi \ell}{d}x_3\Big)} .
    \label{Signal_x3}
\end{eqnarray}
In a symmetric setup with equal grating separations $L$ and periods $d$, the Fourier coefficients are 
\begin{equation}\label{Sell}
    S_\ell =  B_{-\ell}^{(1)}(0) B_{2\ell}^{(2)}\Big(\ell\frac{L}{L_T}\Big) B_{\ell}^{(3)}(0),
\end{equation}
where the Talbot-Lau coefficients $B_{\ell}^{(j)}$ of order $\ell$ for the $j$th grating still need to be determined as a function of the Talbot length $L_T=m v_z d^2/h$.

We assume that every absorbed grating photon results in the ionization of the sodium cluster. The transmission of the particle beam through a standing wave of incident laser power $P$ and Gaussian beam waist $w_y$ is then characterized by the mean number of ionizing photons absorbed in each grating antinode  
\begin{equation}
    n_0 = \frac{8 \sigma_{\rm ion,266} P \lambda_L}{\sqrt{2\pi} h c w_y v_z},
\end{equation}
as well as by the phase shift induced by the optical dipole potential  
\begin{equation}
    \phi_0=\sqrt{\frac{8}{\pi}} \frac{\alpha_{266} P}{\hbar c \varepsilon_0 w_y v_z}.
\end{equation}
The values of the UV polarizability $\alpha_{266}$ and ionization cross section $\sigma_{\rm ion,266}$ are mass-dependent and determined further below. We can then express the Talbot-Lau coefficients as \cite{Nimmrichter2011}
\begin{eqnarray}\label{Bndef}
     B_n(\xi) &=& e^{-n_0/2} \left( \frac{\zeta_{\mathrm{coh}}-\zeta_{\mathrm{ion}}}{\zeta_{\mathrm{coh}}+\zeta_{\mathrm{ion}}} \right)^{n/2}\\
     &\times &J_n\left( \mathrm{sgn}(\zeta_{\mathrm{coh}}+\zeta_{\mathrm{ion}}) \sqrt{\zeta_{\mathrm{coh}}^2-\zeta_{\mathrm{ion}}^2} \right)
     \nonumber,
\end{eqnarray}
where the coherent phase shift and the ionization depletion are described by
\begin{eqnarray}\label{zetacoh}
\zeta_{\mathrm{coh}}(\xi) &= &\phi_0 \sin{(\pi \xi)}\\
\zeta_{\mathrm{ion}}(\xi) &= &\frac{n_0}{2} \cos{(\pi \xi)}.
\label{zetaion}
\end{eqnarray}
For short de Broglie wavelengths,  as  $\xi\equiv L/L_T\to0$,  the latter turn asymptotically into the expressions 
\begin{eqnarray}\label{zetacohcl}
\zeta_{\mathrm{coh}}^{\mathrm{cl}}(\xi) &=& \phi_0 \pi \xi\\
\zeta_{\mathrm{ion}}^{\mathrm{cl}} &=& {n_0}/{2},
\label{zetaioncl}
\end{eqnarray}
which appear in the classical description. It yields the same expression (\ref{Sell})-(\ref{Bndef}) for the signal, except that (\ref{zetacoh}),(\ref{zetaion}) are replaced by (\ref{zetacohcl}),(\ref{zetaioncl}).

In our setup, both the quantum and the classical signal are well approximated by a sinusoidal with  fringe visibility  $V=2|S_1|/S_0$. We average the predicted signal over the measured velocity and mass distributions, accounting for the mass dependence of both the polarizability and the ionization cross section.

\subsection*{Macroscopicity assessment}

To assess the macroscopicity of the demonstrated quantum superposition, it is necessary to calculate how the predicted interference signal is affected by the class of minimal macrorealist modifications (MMM) of quantum mechanics \cite{Nimmrichter2013}.
These are parameterized by the classicalization time scale $\tau_e$, and by the momentum  spread $\sigma_q$ and spatial spread $\sigma_s$ of a phase space distribution.  The greater the value of $\tau_e$, the larger the scales at which the quantum superposition principle still holds.

For our symmetric Talbot-Lau setup the impact of an MMM is accounted for by multiplying the Fourier coefficients (\ref{Sell}) by 
\begin{eqnarray}\label{Rell}
    R_\ell &=&
    \exp\Big[
    -2\sqrt{\frac{2}{\pi}}
    \Big(\frac{3\hbar m}{R_{\rm cl}\sigma_q m_e}\Big)^2
    \frac{L}{v_z \tau_e}
    \\
    &&\times 
    \int_0^\infty     {\rm d}z\,
    {\rm e}^{-z^2/2}
    j_1^2\Big(\frac{R\sigma_q}{\hbar}z\Big)
    f\Big(\frac{\ell d \sigma_q L}{\hbar L_T} z\Big)
    \Big]
    \nonumber
\end{eqnarray}
with $R_{\rm cl}$ the radius of the spherical clusters, $m_e$ the electron mass, $j_1$ a spherical Bessel function, and $f(x) = 1-{\rm Si}(x)/x$ involving the sine integral \cite{Nimmrichter2013}. The dependence on $\sigma_s$ can be neglected for this setup.
The mean count rate is unaffected by MMM since $R_0=1$.

The macroscopicity is obtained by using the raw experimental data $\mathcal{C}$ (cluster counts at given grating shift $x_3$ and grating powers) for a Bayesian test of the hypothesis that MMM holds with a classicalization time no greater than $\tau_e$ \cite{Schrinski2019}.
Bayesian updating yields the posterior probability distribution $p(\tau_e |\mathcal{C}, \sigma_q)$ of the classicalization time $\tau_e$, starting from Jeffreys' prior, by using the likelihoods obtained by incorporating  (\ref{Rell}) in the detection probability $S(x_3)$, see
\cite{Schrinski2020}. The lowest 5\% quantile $\tau_m (\sigma )$ of the posterior distribution
then determines the macroscopicity as
$\mu=\max_{\sigma_q}\log_{10}(\tau_m(\sigma)/1\rm s)$.

In our case, a total number of 3895 data points yield a distribution very well approximated by a Gaussian (Kullback-Leibler divergence $1.27\times 10^{-3}$) whose 5\% quantile $\tau_m=2.84\times 10^{15}$\,s (maximized at $\hbar/\sigma_q=10$\,nm) remains constant to three decimal places after 3280 data points. This indicates that sufficient data was recorded and that the distribution  is independent of the prior. The resulting macroscopicity is $\mu=15.45$.

\subsection*{Cluster beam}
Large sodium clusters are generated in a home-built aggregation chamber, inspired by earlier work \cite{Haberland1991,Goehlich1991}. The sodium is evaporated at $650-700$\,K into a cold mixture of argon and helium at a temperature of $80$\,K and pressure of less than $ 1\,$\,mbar. The resulting distribution covers masses  beyond 1\,MDa and velocities between $120 - 170$\,m/s. The clusters exit through a 5\,mm aperture and pass three differential pumping stages before they reach the interferometer (see Suppl. Inf.)

Two horizontal collimation slits $d_{\text{H1,H2}}=0.5$\,mm spaced by $1.8$\,m facilitate the alignment of the grating yaw angles perpendicular to the molecular beam with a precision of about $200\,\mu$rad. Two vertical collimation slits $d_{\text{V1}}=0.5$\,mm and $d_{\text{V2}}=1$\,mm, spaced by $2.2$\,m, confine the beam height and ensure good overlap with the standing light wave. This also reduces the influence of gravitationally induced phase averaging.

\subsection*{Photophysics}

The optical polarizability $\alpha_\mathrm{266}$, absorption cross-section $\sigma_\mathrm{266}$, and ionization potential $E_i$ depend on the size, mass, and purity of the cluster. They determine transmission, the maximal matter-wave phase shift $\phi_0$ and the mean number of absorbed photons $n_0$ in the antinodes of the grating.
Photophysics \cite{Wrigge2002} and thermodynamics \cite{Schmidt2001} of small sodium clusters have been extensively studied, and the preparation of particles up to 1\,MDa has been demonstrated before \cite{Goehlich1991}. However, the mass-selected UV polarizability has not been known. Here, we use the high contrast fringe patterns of clusters between $0.4-1$\,MDa to determine it in a mass range where the classical and quantum models predict the same visibilities. We derive a value of $\alpha_\mathrm{266}/\text{atom}  = -4\pi \varepsilon_0 \times (4.5\pm0.5) \,\text{\AA}^3$ (see Suppl. Inf.) which is consistent with the experiments and the quantum model for $m=100-200$\,kDa.

The photo-ionization cross section  $\sigma_\mathrm{ion,266}$ is a product of the absorption cross section $\sigma_\mathrm{abs,266}$ and the ionization yield. It determines the total transmission through the interferometer and influences the highest possible interference contrast. By measuring the mass-selected transmission of the interferometer for different grating powers, we determine an effective cross section of $\sigma_\mathrm{ion,266}=(0.537\times m\text{/kDa} - 1.5)\times 10^{-20}\,\mathrm{m}^2$ for our clusters.

\subsection*{Mass selection and detection}

After passing all gratings, the cluster beam is photo-ionized using  425\,nm light and the cations are filtered by their $m/z$ ratio using a quadrupole mass spectrometer (QMS). The mass filter includes guiding ion optics (Extrel) and $300$\,mm long quadrupole rods (Oxford Applied Research) with a diameter of $25.4$\,mm. The mass filter is operated at a resolution of $\Delta m/m =  0.32$. Interference scans centered on mass $m$ therefore involve clusters within a mass range of $\pm \Delta m/2$, where the transmission function is close to rectangular and taken into account in our models. 
The mass filter was centered at $170$\,kDa. The underlying mass distribution, convoluted with the trapezoidal transmission, shifts the effective mass center towards $172$\,kDa. 

The selected cluster ions are counted by a channel electron multiplier with a conversion dynode at 10\,kV. Electronic dark counts range from $15-100$\,counts per second.

 We must also account for the mixing of multiply charged ions with identical $m/z$ ratios. Based on the measured work function of $W = (2.4 \pm 0.1)$\,eV (Suppl. Inf.), neutral clusters with a diameter of $d_{Cl} \sim 8$\,nm exhibit an ionization threshold of $E_i = 2.53$\,eV, followed by $E_{i,+1} = 2.88$\,eV and $E_{i,+2} = 3.23$\,eV for subsequent ionization processes. The detection laser has a photon energy of $E_{\text{ph}} = 2.92$\,eV and can generate doubly charged ions while triply charged ions remain energetically out of reach. 

We have selected doubly charged clusters in the detector and verified the correct cluster mass by analyzing mass spectra at both low and high detection laser powers (Suppl. Inf.). In the antinodes of the gratings, the 266\,nm light can also lead to multiply charged ions. However, this does not affect the interference pattern, because every ion is removed from the cluster beam via electrostatic deflection, independent of its charge state. Only clusters that remain neutral while passing through all gratings contribute to the final interference pattern.

\subsection*{Velocity Distribution}
The cluster velocity distribution is determined from a time-of-flight measurement, where we imprint a start time signal onto the cluster beam by UV photodepletion close to G$_1$ and we measure the cluster arrival time behind the ionizing mass spectrometer. 
The time-of-flight data is denoised with an adaptive Gaussian low-pass filter and corrected  for the drift time inside the quadrupole where it is slightly accelerated by the entrance voltage $U$ to $v' = v + \sqrt{2eU/m}$.
A convolution of a Gaussian drift time distribution and a rectangular chopper opening function is then fitted to the corrected and denoised data. The results are converted to a velocity distribution. We determine the average velocity and the width of the distribution from the standard deviation of the Gaussian fit. 

Small variations of the mean velocity depend on the gas flow and the particle mass, and the $1\sigma$ width is  $\Delta v/v = 5-7\ \%$. Time of flight and velocity spectra for $m/q=100$\,kTh clusters are shown in the Suppl. Information. 


\subsection*{Deep ultraviolet gratings}
Up to $4$\,W of 532 nm light (Coherent Verdi V18) is converted to up to $1$\,W of 266\,nm UV light via intracavity second harmonic generation (Sirah Wavetrain 2). The UV output is vertically expanded and split into three grating beams, using polarizing beam splitters and  half-wave plates to regulate the power for each grating.  
Cylindrical lenses ($f = 140$\,mm) focus the laser horizontally onto high-reflectivity ($R=99.5\%$) mirrors in vacuum to generate the standing light waves. We have observed power losses of up to 60\% due to the degradation of optical components. The beam waists before the lenses are $W_{1} \times H_{1} =  1130 \times 620$ \textmu m$^2$, $W_{2} \times H_{2} = 1020  \times 575$\,\textmu m$^2$, and $W_{3} \times H_{3} = 1020  \times 575$\,\textmu m$^2$, with $\Delta H_i = \Delta w_i = \pm 50$\,\textmu m. Here, $W_i$ represents the $1/e^2$ waist radii along the molecular beam direction, and $H_i$ is the vertical waist.
At the focus, the Gaussian beam waist is $20$ \textmu m.  This small waist alleviates the alignment requirements with regard to  the cluster beam tilt angle. The waist is still sufficiently large to ensure that the Rayleigh length, $z_R = 4.7$\,mm, is an order of magnitude larger than  the cluster beam width of $500$\,\textmu m.

\subsection*{Interferometer alignment}
The grating mirror surface is aligned parallel to the particle beam axis, with the standing light wave along the mirror normal. The mirror exhibits three angular degrees of freedom: pitch, yaw, and roll. The yaw angle, between  mirror surface and particle beam, is adjusted to better than $200$\,µrad. The mirror roll angles, around the molecular beam axis, are aligned to each other to less than $ 20$\,µrad  and they are stabilized with respect to the Earth's gravitational field to better than $50$\,µrad. 
The distances between the gratings are equal within 50\,µm.

\subsection*{Interference scans}
We obtain the interference scans by measuring the number of transmitted clusters as a function of the transverse displacement of the third grating G$_3$ , which is moved in steps of $\Delta x=15$\,nm. At each position, the mass-filtered ion signal is integrated for a time interval of up to four seconds. A sinusoidal fit to the data then provides the periodicity, phase, and amplitude of the fringes. By design of first-order Talbot-Lau interferometry, the periodicity is equal to the grating period.  

\section*{Declarations}
This work was supported in part by the Gordon and Betty Moore Foundation through Grant 10771 and by the Austrian Science Fund (FWF) Grant 32542-N.
The authors declare no conflict of interest regarding this study. Ethics approval and consent to participate were not applicable, as the study did not involve human participants or animals, and consent for publication is similarly not applicable. Data supporting the findings of this research are available upon reasonable request from the corresponding author, and materials used in the study can also be provided upon request, subject to availability. The code developed and utilized for this research is available upon request to ensure transparency and reproducibility. Author contributions are as follows: SP was responsible for experiment design, measurement, data analysis, theory, and writing; BR and RF were involved in experiment design, measurement, and theory; KH provided theory and the Bayesian estimate of quantum macroscopicity and contributed to writing; MA contributed to experimental design, theory, and writing; SG was responsible for experimental design, analysis, and theory.

\begin{acknowledgments}
We thank P. Geyer, Y.Y. Fein for contributions in the early stage of the experiment, and S. Sindelar for his support with metal cluster beams. MA is indebted to B. v. Issendorff for many discussions on cluster science throughout the years leading to this experiment. The authors thank V. Kresin for comments on alkali cluster sources and A. Shekhoon for discussions on the cluster work function.
\end{acknowledgments}

%

\end{document}